\documentclass[aps,preprintnumbers,superscriptaddress,showpacs]{revtex4}
\usepackage{epsfig}
\usepackage{psfrag}
\usepackage{amsfonts}
\usepackage{graphicx}
\usepackage{dcolumn}
\usepackage{bm}

\begin{document}

\title{Drag force in strongly coupled $\mathcal{N}=4$ SYM plasma in a magnetic field}

\author{Zi-qiang Zhang}
\email{zhangzq@cug.edu.cn} \affiliation{School of mathematics and
physics, China University of Geosciences(Wuhan), Wuhan 430074,
China}

\author{Ke Ma}
\email{make@cug.edu.cn} \affiliation{School of mathematics and
physics, China University of Geosciences(Wuhan), Wuhan 430074,
China}

\author{De-fu Hou}
\email{houdf@mail.ccnu.edu.cn} \affiliation{Key Laboratory of
Quark and Lepton Physics (MOE), Central China Normal University,
Wuhan 430079, China}

\begin{abstract}
Applying AdS/CFT correspondence, we study the effect of a constant
magnetic field $\mathcal{B}$ on the drag force associated with a
heavy quark moving through a strongly-coupled $\mathcal{N}=4$
supersymmetric Yang-Mills (SYM) plasma. The quark is considered
moving transverse and parallel to $\mathcal{B}$, respectively. It
is shown that for transverse case, the drag force is linearly
dependent on $\mathcal{B}$ in all regions. While for parallel
case, the drag force increases monotonously with increasing
$\mathcal{B}$ and also reveals a linear behavior in the regions of
strong $\mathcal{B}$. In addition, we find that $\mathcal{B}$ has
important effect for transverse case than parallel.

\end{abstract}

\maketitle
\section{Introduction}
The experiments at Relativistic Heavy Ion Collider (RHIC) and
Large Hadron Collider (LHC) have produced a new state of matter
so-called quark gluon plasma (QGP) \cite{JA,KA,EV}. One of the
interesting properties of QGP is jet quenching: due to the
interaction with the medium, high energy partons propagating
through the QGP are strongly quenched. Usually, this phenomenon
can be characterized by the jet quenching parameter which
describes the average transverse momentum square transferred from
the traversing parton, per unit mean free path. Alternately, the
energy loss can also be analyzed from the drag force, which is
related to the interaction between the moving quark and the
medium. In the framework of weakly theories, the calculation of
the jet quenching has been studied in many papers, see e.g.
\cite{XN,BG,RB1,RB2,UA,GLV,SJ,AMY}. However, many experimental
results indicate that QGP is strongly coupled \cite{EV}. Thus, one
would like to study the jet quenching in strongly coupled theory
via the use of non-perturbative techniques, such as AdS/CFT
\cite{Maldacena:1997re,Gubser:1998bc,MadalcenaReview}.

AdS/CFT, the duality between the type IIB superstring theory
formulated on AdS$_5\times S^5$ and $\mathcal N=4$ SYM theory in
four dimensions, has yielded many important insights for studying
different aspects of QGP \cite{JCA}. In this approach, the drag
force for $\mathcal{N}=4$ SYM plasma was first investigated in
\cite{CP,GB}. Therein, the energy loss of heavy quark is
understood as the momentum flow along the string into the horizon.
Later, this idea has been extended to various cases. For example,
the effects of chemical potential on the drag force have been
studied in \cite{ECA,LC}. The $R^2$ correction on the drag force
have been discussed in \cite{KBF}. The effects of constant B-field
or non-commutativity on drag force have been investigated in
\cite{TMA}. Also, for the drag force in STU background, see
\cite{JSA}. For this quantity in some AdS/QCD models, see
\cite{ENA,PE,UGU}. Other important results can be found, for
example, in \cite{DGI,SCH,MCH,ANA,KLP,SRO,SSG1,NA}.

Now we would like to give such analysis under the influence of a
magnetic field. The motivation comes from the experiment: the QGP
produced in heavy-ion collisions may be subject to a strong
electromagnetic field that created by many spectator nucleons
\cite{DEK0} and the effect of a magnetic field on some topological
\cite{DT,KF,DEK1} and the dynamical \cite{GSB,KA1,DD,RR}
properties of QGP have been investigated recently. On the other
hand, one would like to be able to use holography to study the
effect of the magnetic field on various quantities
\cite{ED,RCR,KAMA,RRO,KIM,SI,SLI}. Not long ago, the drag force in
a strongly coupled $\mathcal{N}=4$ SYM plasma with a strong
magnetic field has been discussed in \cite{KIM} and the results
show that this force is linearly dependent on $\mathcal{B}$, i.e.,
$f=-\frac{\sqrt{\lambda}\mathcal{B}}{6\pi}v$. However, the metric
therein is valid only near the horizon, so the discussions are
restricted to the infrared (IR) regime. In this paper, we would
like to extend it to the case of all regimes by considering a
general magnetic field. Specially, we want to know how an
arbitrary magnetic field affects the drag force.

The paper is organized as follows. In the next section, we briefly
review the asymptotic $AdS_5$ holographic Einstein-Maxwell model
and introduce the background metric in the presence of a magnetic
field. In section 3, we show numerical procedure and some
numerical solutions. In section 4, we investigate the drag force
for the quark moving transverse and parallel to the magnetic
field, in turn. The last part is devoted to conclusion and
discussion.


\section{Background geometry}
The holographic model is Einstein gravity coupled with a Maxwell
field, corresponding to strongly coupled $\mathcal{N}=4$ SYM
subjected to a constant and homogenous magnetic field. The bulk
action is \cite{ED}
\begin{equation}
S=\frac{1}{16\pi G_5}\int
d^5x\sqrt{-g}(R+\frac{12}{L^2}-F^2)+S_{body},\label{action}
\end{equation}
where $G_5$ is the 5-dimensional gravitational constant, $L$
denotes the radius of the asymptotic $AdS_5$ spacetime. $F$ stands
for the Maxwell field strength 2-form. Moreover, the term
$S_{body}$ contains the Chern-Simons terms, Gibbons-Hawking terms
and other contributions necessary for a well posed variational
principle, but $S_{body}$ does not affect the solutions considered
here.

The equations of motion for (\ref{action}) are given by the
Einstein equations
\begin{equation}
R_{\mu\nu}+\frac{4}{L^2}g_{\mu\nu}+\frac{1}{3}F_{\rho\sigma}F^{\rho\sigma}g_{\mu\nu}-2F_{\mu\rho}F_\nu^\rho=0,
\label{ein}
\end{equation}
and the Maxwell's field equations
\begin{equation}
\nabla_\mu F^{\mu\nu}=0.
\end{equation}

The ansatz for the magnetic brane geometry is \cite{ED}
\begin{equation}
ds^2=-H(r)dt^2+e^{2P(r)}(dx^2+dy^2)+e^{2V(r)}dz^2+\frac{dr^2}{H(r)},\label{metric}
\end{equation}
with
\begin{equation}
F=Bdx\wedge dy,
\end{equation}
where, for simplicity, we have set $L=1$. Note that in
(\ref{metric}) the horizon is located at $r=r_h$ with $H(r_h)=0$.
The boundary is located at $r=\infty$. The constant $B$ refers to
the bulk magnetic field, pointing in the $z$ direction. Also, the
three coefficients $H(r)$, $P(r)$, and $V(r)$ can be obtained by
solving the equations of motion.

In terms of (\ref{metric}), the Einstein equations reduce to
\begin{equation}
H(P^{\prime\prime}-V^{\prime\prime})+(H^\prime+H(2P^\prime+V^\prime))(P^\prime-V^\prime)=-2B^2e^{-4P}\label{e1},
\end{equation}
\begin{equation}
2P^{\prime\prime}+V^{\prime\prime}+2(P^\prime)^2+(V^\prime)^2=0,\label{e2}
\end{equation}
\begin{equation}
\frac{1}{2}H^{\prime\prime}+\frac{1}{2}H^{\prime}(2P^\prime+V^\prime)=4+\frac{2}{3}B^2e^{-4P},\label{e3}
\end{equation}
\begin{equation}
2H^\prime P^\prime+H^\prime V^\prime+2H(P^\prime)^2+4HP^\prime
V^\prime=12-2B^2e^{-4P},\label{e4}
\end{equation}
where the derivations are with respect to $r$. Unfortunately, for
these coupled equations analytic solution can not be obtained
easily. But an exact solution near the horizon ($r\sim r_h$),
which denotes the product of a Banados, Teitelboim and Zanelli
(BTZ) black hole times a two dimensional torus $T^2$, can be found
as
\begin{equation}
ds^2=-\frac{r^2f(r)}{\mathcal{R}^2}dt^2+\mathcal{R}^2\mathcal{B}(dx^2+dy^2)+\frac{r^2}{\mathcal{R}^2}dz^2+\frac{\mathcal{R}^2}{r^2f(r)}dr^2,\label{metric1}
\end{equation}
with $f(r)=1-\frac{r_h^2}{r^2}$. Here $\mathcal{B}=\sqrt{3}B$
represents the physical magnetic field at the boundary and
$\mathcal{R}=\frac{L}{\sqrt{3}}=\frac{1}{\sqrt{3}}$ denotes the
radius of the BTZ black hole. It should be emphasized that the
metric (\ref{metric1}) is only valid near the horizon, i.e., in
the regime $r<<\sqrt{\mathcal{B}}\mathcal{R}^2$ where the scale is
much smaller than the magnetic field. Recently, several authors
have used this metric to study the effect of strong magnetic field
(IR regime) on the energy loss \cite{KIM} and jet quenching
parameter \cite{SLI}.

In this article, rather than using (\ref{metric1}), we apply a
solution that interpolates between (\ref{metric1}) in the IR and
$AdS_5$ in the ultraviolet (UV). From the point of view of the
boundary theory, this refers to an renormalization group (RG) flow
between a $D=3+1$ CFT at small $r$ and a $D=1+1$ CFT at large $r$
\cite{ED}. However, no analytic solution can be found in this case
and one needs to resort to numerics. In the next section, we
follow the numerical procedure mentioned in \cite{ED} and present
some numerical solutions.

\section{Numerical solutions}
To begin with, we derive some useful equations. By eliminating the
$B^2e^{-4P}$ terms in (\ref{e1})$-$(\ref{e4}), we have
\begin{equation}
3H^{\prime\prime}+5(V^\prime+2P^\prime)H^\prime+4({P^\prime}^2+2P^\prime
V^\prime)H-48=0,\label{e5}
\end{equation}
\begin{equation}
3HP^{\prime\prime}+2H{P^\prime}^2-H^\prime P^\prime-5HP^\prime
V^\prime+12-2H^\prime V^\prime=0,\label{e6}
\end{equation}
\begin{equation}
3HV^{\prime\prime}+3H{V^\prime}^2+4H^\prime V^\prime+10HP^\prime
V^\prime+2H{P^\prime}^2+2H^\prime P^\prime-24=0.\label{e7}
\end{equation}

Following \cite{ED}, it is convenient to use rescaled coordinates.
First, we rescale $t\rightarrow\bar{t}$ , $r\rightarrow\bar{r}$
and fix the horizon at $\bar{r}_h=1$, so that
\begin{equation}
H(1)=0,\qquad H^{\prime}(1)=1. \label{h}
\end{equation}

In this case, the Hawking temperature is
\begin{equation}
T=\frac{\sqrt{-g^\prime_{\bar{t}\bar{t}}g^{\bar{r}\bar{r}\prime}}}{4\pi}\big{|}
_{\bar{r}=1}=\frac{1}{4\pi}.
\end{equation}

Next we rescale $x$, $y$, $z$ coordinates to have
\begin{equation}
P(1)=V(1)=0,\qquad P^\prime(1)=4-\frac{b^2}{3},\qquad
V^\prime(1)=4+\frac{b^2}{6}, \label{w}
\end{equation}
where $b$ stands for the value of the magnetic field in the
rescaled coordinates. Notice that if $P^\prime<0$, the geometry
will not be asymptotically $AdS_5$, thus, the second equation in
(\ref{w}) gives us $0\leq b<2\sqrt{3}$.

Moreover, the geometry will have the asymptotic behavior as
$\bar{r}\rightarrow \infty$,
\begin{equation}
H(\bar{r})\rightarrow \bar{r}^2, \qquad e^{2P(\bar{r})}\rightarrow
m(b) \bar{r}^2, \qquad e^{2V(\bar{r})}\rightarrow n(b) \bar{r}^2,
\end{equation}
where $m(b)$ and $n(b)$ are rescaling parameters which can be
determined numerically. Also, the physical magnetic field $B_0$ is
given by
\begin{equation}
B_0=\sqrt{3}\frac{b}{m(b)}.\label{b}
\end{equation}

Actually, the interval of $b$ can be obtained from (\ref{b}) as
well. One can numerically check that $m(b)$ is a decreasing
function of $b$ and $m(b\rightarrow 2\sqrt{3})\rightarrow 0$, for
this behavior, see in the left panel of fig.1. On the other hand,
it implies that one covers in practice all values of $B_0$ for
$0\leq b<2\sqrt{3}$.

Finally, to have an asymptotic $AdS_5$ in the UV, one needs to
rescale back to the original coordinate system by setting
$(\bar{x},\bar{y},\bar{z})\rightarrow
(x/\sqrt{m(b)},y/\sqrt{m(b)},z/\sqrt{n(b)}$, then metric
(\ref{metric}) becomes
\begin{equation}
ds^2=-H(\bar{r})dt^2+\frac{e^{2P(\bar{r})}}{m(b)}(dx^2+dy^2)+\frac{e^{2V(\bar{r})}}{n(b)}dz^2+\frac{dr^2}{H(\bar{r})}.
\end{equation}

When this is done, one can solve the coupled equations
(\ref{e5})$-$(\ref{e7}) with the boundary conditions (\ref{h}) and
(\ref{w}). As a matter of convenience, we drop from now on the
bars in the rescaled coordinates. The numerical procedure can be
summarized as follows: First, choosing a value of $b$ for $0\leq
b<2\sqrt{3}$, one solves the equations (\ref{e5})$-$(\ref{e7}) and
obtains the numerical solutions of $H(r), P(r)$ and $V(r)$. Then,
fitting the asymptotic data for $e^{2P(r)}\rightarrow m(b) r^2$
and $e^{2V(r)}\rightarrow n(b) r^2$, one gets the values of $m(b)$
and $n(b)$. Then the value of $B_0$ can be obtained from
(\ref{b}). Finally, setting $(e^{2P(r)},e^{2V(r)})\rightarrow
(e^{2P(r)}/m(b),e^{2V(r)}/n(b))$, one obtains the numerical
solutions. Likewise, one can study other cases by varying the
value of $b$ . In the right panel of fig.1, we plot $\ln H(r),
P(r)$, $V(r)$ versus $r$ for $b=2.7$, we have checked that it
matches the fig.3 in \cite{RRO}.

\begin{figure}
\centering
\includegraphics[width=8cm]{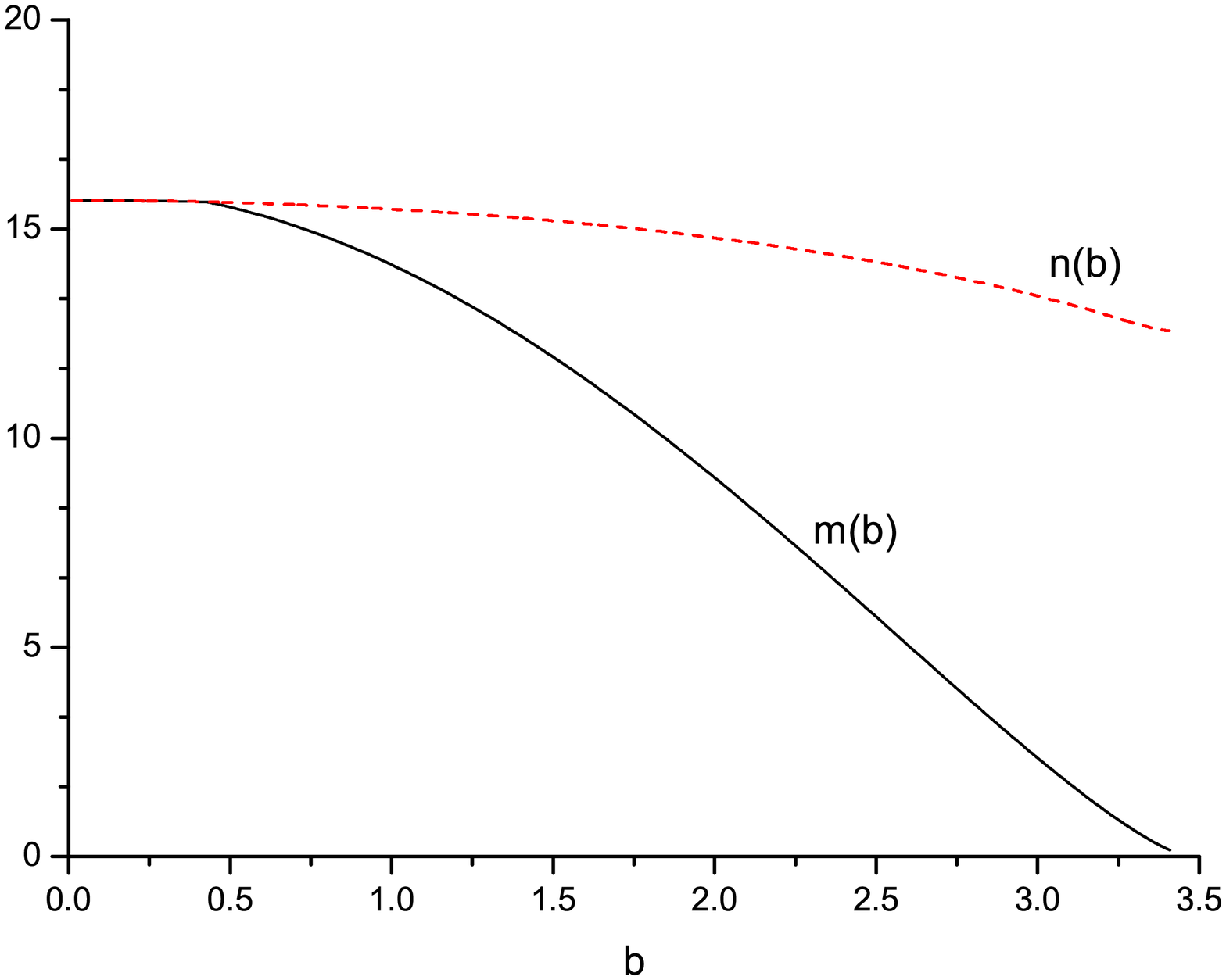}
\includegraphics[width=8cm]{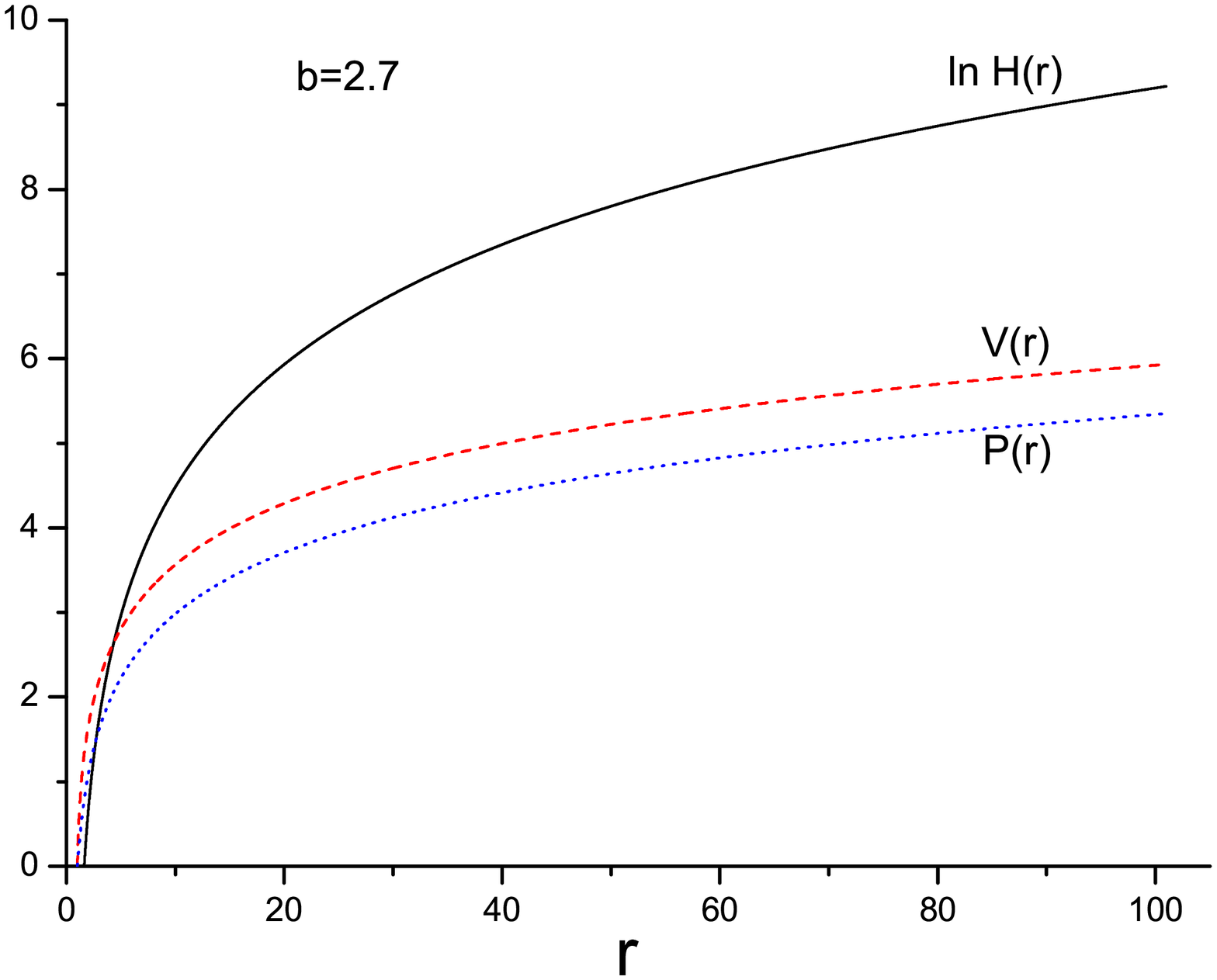}
\caption{Left: $n(b)$ (dash curve) and $m(b)$ (solid curve)
against $b$. Right: $\ln H(r)$ (solid curve), $V(r)$ (dash curve)
and $P(r)$ (dot curve) against $r$ for $b=2.7$.}
\end{figure}

\section{drag force}
It is known that when a heavy quark moves in a hot medium, its
interaction with the medium leads to a drag force thus making it
losing energy. On the other hand, the energy loss can be depicted
in a dual trailing string picture \cite{CP,GB}, that is, a heavy
quark moving on the boundary, but with a string tail into the AdS
bulk. Under this scenario, the dissipation of the heavy quark
could be described by the drag force, which is conjectured to be
associated with a string tail in the fifth dimension.

In the proposal of \cite{CP,GB}, the drag force is related to the
damping rate $\mu$ (or friction coefficient), defined by Langevin
equation,
\begin{equation}
\frac{dp}{dt}=-\mu p+f_1,
\end{equation}
subject to a driving force $f_1$. And, for constant speed
trajectory (or $dp/dt=0$), the driving force is equivalent to a
drag force $f$.

Generally, to discuss the magnetic effect, one needs to consider
different alignments for the velocity with respect to the
direction of the magnetic field, i.e., transverse
($\theta=\pi/2$), parallel ($\theta=0$), or arbitrary direction
($\theta$). Here we consider two cases: transverse and parallel.

\subsection{Transverse case ($\theta=\pi/2$)}
First we study the quark moving perpendicularly to the magnetic
field in the $x$ direction. The coordinates are parameterized by
\begin{equation}
t=\tau, \qquad x=vt+\xi(r),\qquad y=0,\qquad z=0,\qquad
r=\sigma,\label{par}
\end{equation}
where one end point of the trailing string moves with the velocity
$v$ on the boundary while the other parts move in the bulk.

The string dynamic is governed by the Nambu-Goto action
\begin{equation}
S=-\frac{1}{2\pi\alpha^\prime}\int d\tau d\sigma\sqrt{-g},
\label{S}
\end{equation}
where $g$ is the determinant of the induced metric with
\begin{equation}
g_{\alpha\beta}=g_{\mu\nu}\frac{\partial
X^\mu}{\partial\sigma^\alpha} \frac{\partial
X^\nu}{\partial\sigma^\beta},
\end{equation}
where $g_{\mu\nu}$ and $X^\mu$ represent the brane metric and
target space coordinates, respectively. Moreover, $\alpha^\prime$
is related to the 't Hooft coupling by
$1/\alpha^\prime=\sqrt{\lambda}$.

Plugging (\ref{par}) into (\ref{metric}), the induced metric reads
\begin{equation} g_{tt}=-H(r), \qquad
g_{xx}=e^{2P(r)},\qquad g_{rr}=\frac{1}{H(r)},
\end{equation}
given this, one can identify the lagrangian density as
\begin{equation}
\mathcal
L=\sqrt{-g_{rr}g_{tt}-g_{rr}g_{xx}v^2-g_{xx}g_{tt}{\xi^\prime}^2}=\sqrt{1-\frac{e^{2P(r)}}{H(r)}v^2+e^{2P(r)}H(r){\xi^\prime}^2},
\end{equation}
with $\xi^\prime=d\xi/d\sigma$.

The equation of motion implies that $\frac{\partial
L}{\partial\xi^\prime}$ is a constant. If one calls it $\Pi_\xi$,
then
\begin{equation}
\Pi_\xi=\frac{\partial
L}{\partial\xi^\prime}=-\xi^\prime\frac{g_{tt}g_{xx}}{\sqrt{-g}}=\xi^\prime\frac{H(r)e^{2P(r)}}{\sqrt{1-\frac{e^{2P(r)}}{H(r)}v^2+e^{2P(r)}H(r){\xi^\prime}^2}}\label{lag},
\end{equation}
results in
\begin{equation}
{\xi^\prime}^2=\frac{{\Pi_\xi}^2[1-\frac{v^2e^{2P(r)}}{H(r)}]}{H(r)e^{2P(r)}[H(r)e^{2P(r)}-{\Pi_\xi}^2]},\label{xi}
\end{equation}
note that in the right hand of (\ref{xi}), near the horizon the
denominator and numerator are both positive for large $r$ and
negative for small $r$. In addition, it is required that
${\xi^\prime}^2$ must be everywhere positive. Thus, the
denominator and numerator should change sigh at the same point,
which leads to
\begin{equation}
H(r_c)=e^{2P(r_c)}v^2,\label{hrc}
\end{equation}
and
\begin{equation}
{\Pi_\xi}^2=H(r_c)e^{2P(r_c)},
\end{equation}
where $r=r_c$ is the critical point.

On the other hand, the current density for momentum $p_1$ along
the $x$ direction can be written as
\begin{equation}
\pi_x^r=-\frac{1}{2\pi\alpha^\prime}\xi^\prime\frac{g_{tt}g_{xx}}{-g}.
\end{equation}

As a result, the drag force is obtained as
\begin{equation}
f_b=\frac{dp_1}{dt}=\sqrt{-g}\pi_x^r=-\frac{1}{2\pi\alpha^\prime}{\Pi_\xi}=-\frac{1}{2\pi\alpha^\prime}ve^{2P(r_c)},\label{drag}
\end{equation}
where the minus sign means that the direction of the drag force is
against the movement.

To compare with the strong magnetic field case in \cite{KIM}, we
set $e^{2P(r_c)}=\mathcal{R}^2\mathcal{B}=\frac{1}{3}\mathcal{B}$
in (\ref{drag}). After using the relation
$1/\alpha^\prime=\sqrt{\lambda}$, one gets
\begin{equation}
\frac{dp_1}{dt}=-\frac{\sqrt{\lambda}\mathcal{B}}{6\pi}v,
\label{dragm}
\end{equation}
which is exactly Eq.(32) in \cite{KIM}.

Also, if one sets $H(r)=\frac{r^2}{L^2}(1-\frac{r_h^4}{r^4})$,
$e^{2P(r)}=e^{2V(r)}=\frac{r^2}{L^2}$ in (\ref{metric}), the drag
force for $\mathcal{N}=4$ case \cite{CP,GB} can be obtained from
(\ref{drag}), that is
\begin{equation}
f_{\mathcal{N}=4}=-\frac{\pi
T^2\sqrt{\lambda}}{2}\frac{v}{\sqrt{1-v^2}},
\end{equation}
where we have used the relations
\begin{equation}
L^4=\lambda {\alpha^\prime}^2,\qquad r_h=\pi L^2T.
\end{equation}

To proceed, we study the effect of magnetic field on the drag
force for the transverse case. Numerically, we plot the absolute
value of the drag force versus $B_0/T^2$ in the left panel of
fig.2. (Since our main interest is to consider the magnetic field
effect, the coefficient $\frac{1}{2\pi\alpha^\prime}$ does not
play any role, here we set it as unity). From the figures, one can
see that at fixed velocity the drag force is almost linearly
dependent on $B_0/T^2$. Especially, the linear behavior is quite
well for the regions of strong magnetic field, in accordance with
\cite{KIM}. Moreover, by comparing the two figures, one finds that
at fixed magnetic field, the drag force increases as the velocity
increases.

\begin{figure}
\centering
\includegraphics[width=8cm]{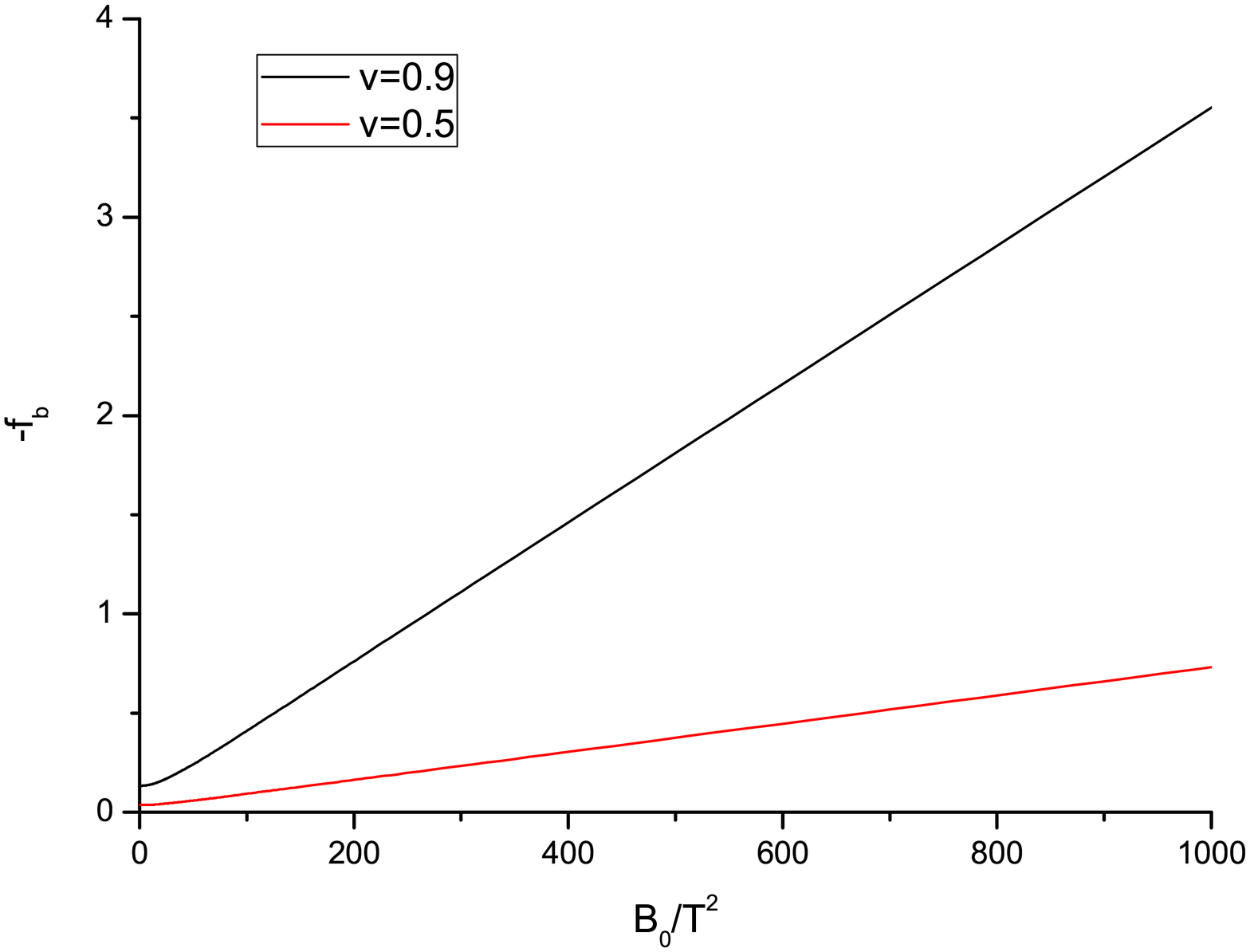}
\includegraphics[width=8cm]{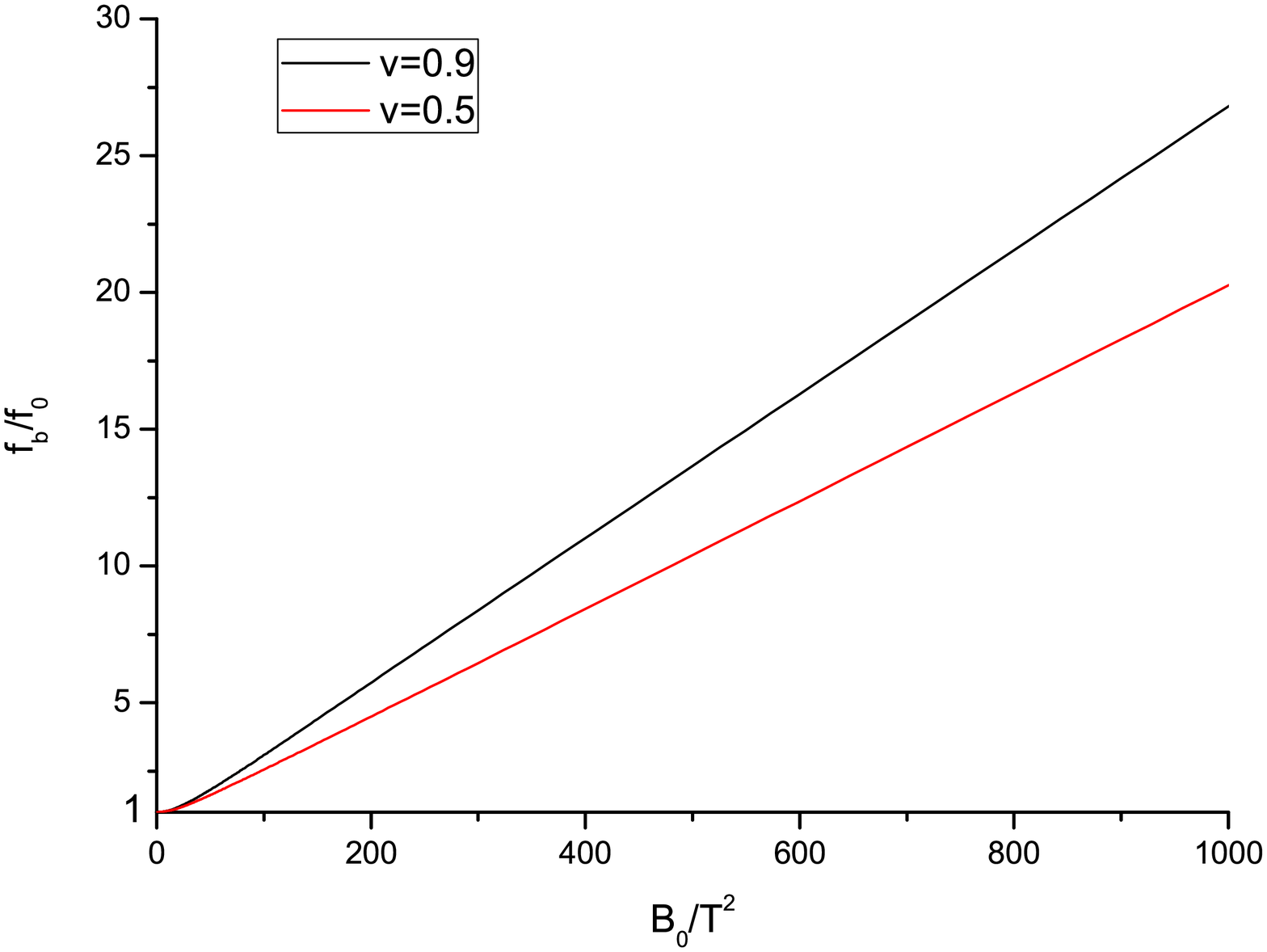}
\caption{Left: ${-f_b}$ against $B_0/T^2$ for $\theta=\pi/2$.
Right: ${f_b}/{f_0}$ against $B_0/T^2$  for $\theta=\pi/2$. In all
of the plots from top to bottom, $v=0.9,0.5$, respectively. Here
the velocity of light is taken as $c=1$.}
\end{figure}

On the other hand, one can compare the drag force between the
cases of $b\neq0$ and $b=0$ as following,
\begin{equation}
\frac{f_b}{f_0}=\frac{(\frac{dp_1}{dt})_{b}}{(\frac{dp_1}{dt})_{b=0}}=\frac{\frac{e^{2P(r_c)}}{m(b)}|_b}{\frac{e^{2P(r_c)}}{m(b)}|_{b=0}},
\end{equation}
the plots of ${f_b}/{f_0}$ versus $B_0/T^2$ for two different
velocities are presented in the right panel of fig.2. One can see
that it also reveals a linear behavior. Therefore, one concludes
that for the transverse case, the drag force increases linearly
with the increase of the magnetic field.

\subsection{Parallel case ($\theta=0$)}
In this subsection we discuss the heavy quark moving parallel to
the magnetic field in the $z$ direction. The coordinates are
parameterized by
\begin{equation}
t=\tau, \qquad x=0,\qquad y=0,\qquad z=vt+\xi(r).\qquad
r=\sigma.\label{par1}
\end{equation}

The next analysis is very similar to the transverse case, so we
present the final results. The drag force for the parallel case is
\begin{equation}
f_b^\prime=\frac{dp_2}{dt}=\sqrt{-g^\prime}\pi_z^r=-\frac{1}{2\pi\alpha^\prime}ve^{2V(r_c)},\label{drag1}
\end{equation}
where $r_c$ satisfies
\begin{equation}
H(r_c)=e^{2V(r_c)}v^2.\label{hrc1}
\end{equation}

Likewise, we plot ${-f_b^\prime}$ versus $B_0/T^2$ and
${f_b^\prime}/{f_0}$ versus $B_0/T^2$ in fig.3. From the figures,
one finds that the drag force monotonously increases as the
magnetic field increases. Also, it reveals a linear behavior for
the regime of strong magnetic field. In addition, by comparing
fig.2 and fig.3, one can see that the slope of the plot in the
parallel case is much less than its counterpart in the transverse
case, which means that the magnetic field has important effect for
the transverse case. Interestingly, a similar observation has been
found in \cite{RRO} which indicates that the magnetic field has
stronger effect on the heavy quark potential for the perpendicular
configuration.

\begin{figure}
\centering
\includegraphics[width=8cm]{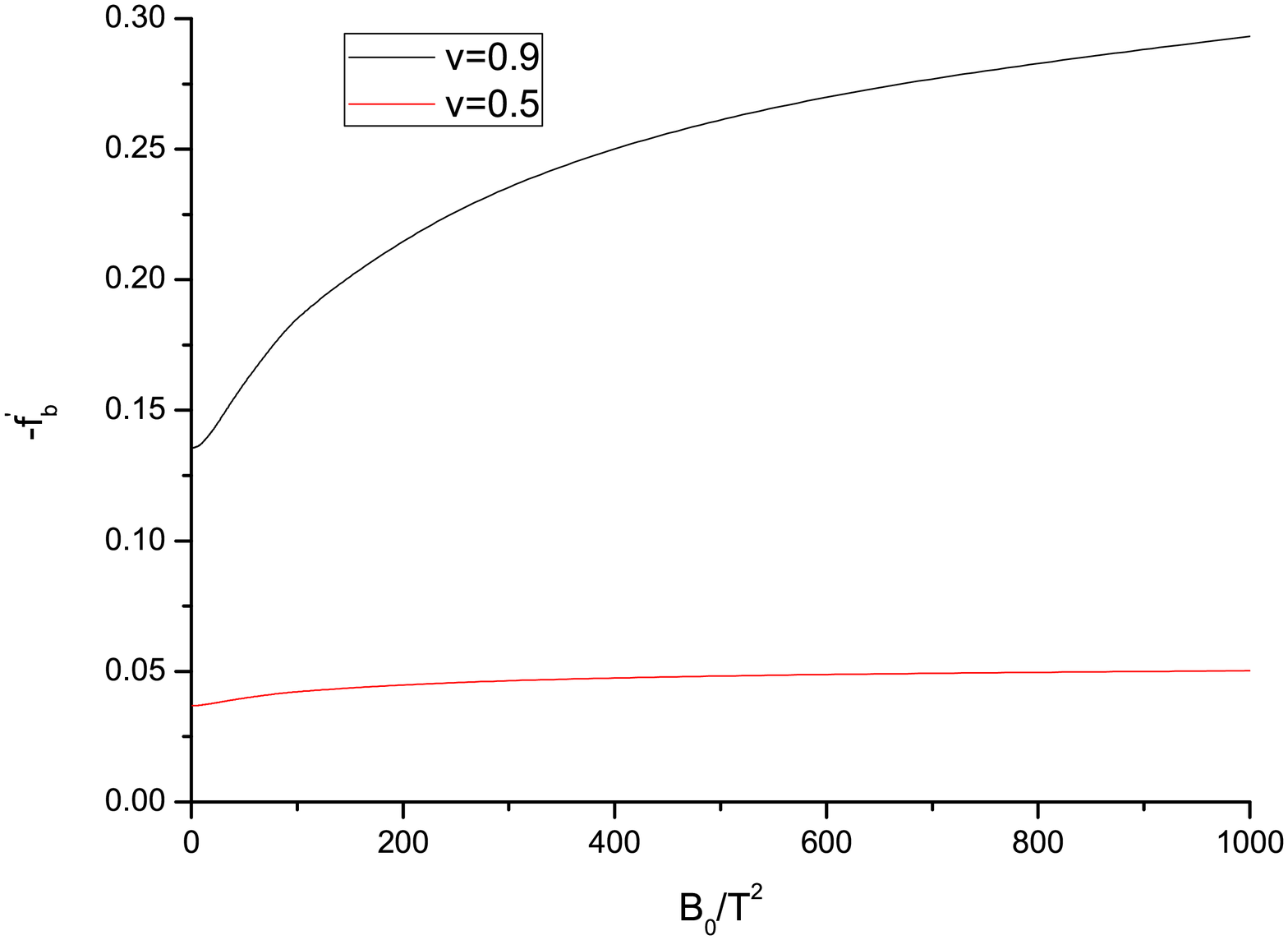}
\includegraphics[width=8cm]{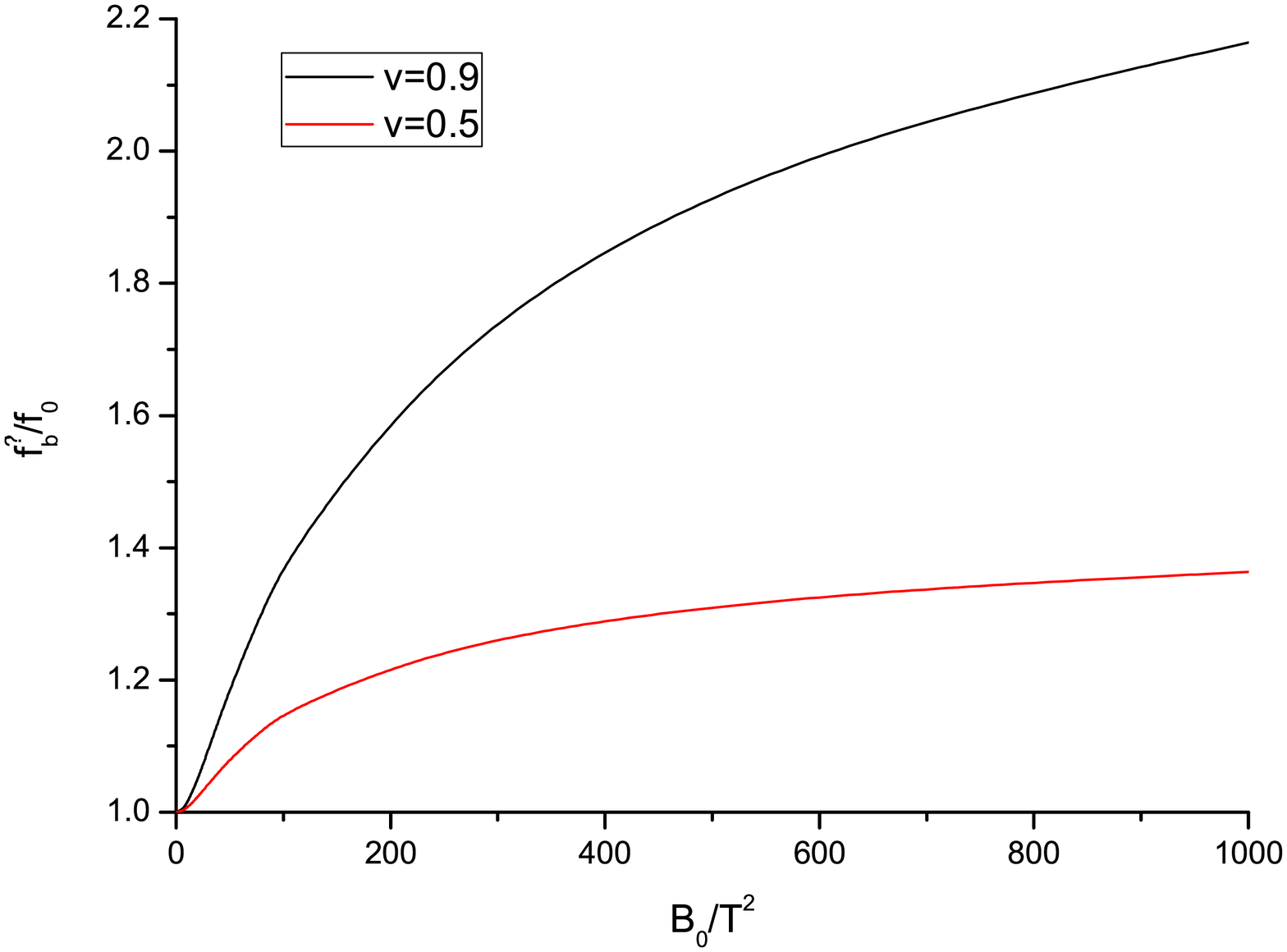}
\caption{Left: ${-f_b^\prime}$ against $B_0/T^2$  for $\theta=0$.
Right: ${f_b^\prime}/{f_0}$ against $B_0/T^2$  for $\theta=0$. In
all of the plots from top to bottom, $v=0.9,0.5$, respectively.
Here the velocity of light is taken as $c=1$.}
\end{figure}

Several comments are in order: First, the rate of energy loss
$\frac{dE}{dt}=\bar{{f}}\cdot \bar{v}$ is dependent on the
magnetic field, and a strong magnetic field also yields a linear
behavior, in agreement with \cite{KIM}. On the other hand, it is
known that the drag force is a kind of viscous force, since the
magnetic field has the effect of increasing the drag force, one
can say that the magnetic field makes the medium more viscous. One
step further, the magnetic field increases the effective viscosity
of QGP to a heavy quark.

\section{conclusion and discussion}
Motivated by the recent studies which regarding the influence of a
strong magnetic field on QGP, in this paper, we analyzed the
effect of a constant magnetic field on the drag force with respect
to a heavy quark moving in a strongly-coupled $\mathcal{N}=4$ SYM
plasma. We considered the quark moving transverse and parallel to
the magnetic field, respectively. It is shown that for transverse
case, the drag force is linearly dependent on the magnetic field.
While for parallel case, the drag force monotonously increases as
the magnetic field increases, and in the regions of strong
magnetic field it also reveals a linearly behavior, which supports
the findings of \cite{KIM}. In addition, we find that the magnetic
field has a stronger effect for transverse case rather than
parallel.

On the other hand, the results indicate that the magnetic field
increases the effective viscosity of QGP. Interestingly, this
finding is contrast to that in \cite{TMA}. But one should keep in
mind that the two results come from two different holographic
models. In \cite{TMA}, the authors consider a non-magnetized
plasma (ignoring the effect of the magnetic field on the plasma)
and discuss the effect of constant B-field or non-commutativity.
In this article, we consider a strongly coupled $\mathcal{N}=4$
SYM plasma in the presence of a constant magnetic field. In short,
the later is closer to practical \cite{KIM}.

Certainly, one should bear in mind that $\mathcal{N}=4$ SYM and
QCD are different theories, in particular, in the vacuum. But at
finite temperature the two theories appear less different. So the
results obtained from $\mathcal{N}=4$ SYM plasma may shed some
light to QGP.

Finally, it should be noticed that the plasma considered here is
with zero chemical potential and finite magnetic field. So one can
take account into finite density in this model as well. It is
relevant to mention that the charged magnetic brane solutions has
been discussed in \cite{ED1}. Using that metric, one can study the
effects of both chemical potential and finite magnetic field on
the drag force. We leave this for further study.

\section{Acknowledgments}
The authors would like to thank the anonymous referee for his/her
valuable comments and helpful advice. This work is partly
supported by the Ministry of Science and Technology of China
(MSTC) under the ¡°973¡± Project No. 2015CB856904(4). Z-q Zhang is
supported by NSFC under Grant No. 11705166. D-f. Hou is supported
by the NSFC under Grants Nos. 11735007, 11521064.


\end{document}